# A Cyber-Physical System-based Approach for Industrial Automation Systems

Kleanthis Thramboulidis
Electrical and Computer Engineering
University of Patras, Greece

***Abstract*—Industrial automation systems (IASs) are commonly developed using the languages defined by the IEC 61131 standard and are executed on PLCs. Their software part is considered after the development and integration of mechanics and electronics. This approach narrows the solution space for software and is considered inadequate to address the complexity of today's systems. In this paper, a system-based approach for the development of IASs is adopted. A framework is described to refine the UML model of the software part, which is extracted from the SysML system model, and get the implementation code. Two implementation alternatives are considered to exploit PLCs but also the recent deluge of embedded boards in the market. For PLC targets, the new version of IEC 61131 that supports Object-Orientation is adopted, while Java is used for embedded boards. The case study was developed as a lab exercise for teaching the various technologies that address challenges in the domain of cyber-physical systems where Internet of Things (IoT ) would be the glue regarding their cyber interfaces**

## I. INTRODUCTION

INDUSTRIAL automation systems (IASs) are composed of the physical plant, which performs the physical processes, and networks of embedded computers, which perform the computational processes required to monitor and control the physical ones. Computational processes, which constitute the cyber part of the system, accept inputs from the physical processes, calculate the outputs required to affect the physical processes and apply these outputs to the physical plant, i.e., the physical part of the system. This is usually realized using time triggered control in the form of the well known scan cycle paradigm.

Computational processes are commonly implemented based on the de-facto standard IEC 61131, which defines a set of languages for programming on PLCs [1]. This 20 years old standard has introduced in the industrial automation domain basic concepts of object orientation through the construct of Function Block (FB) [2]. However, as the complexity of IASs increases and flexibility is of higher priority, the 20 years old technology is not able to address the new requirements [3].

To address the restrictions imposed by version 2.0 of IEC 61131, as well as to address the new challenges in the development of today's complex industrial automation systems, the IEC has defined the IEC 61499 standard [4]. This standard "has emerged in response to the technological limitations encountered in the currently dominating standard IEC 61131", as claimed in [5], where IEC 61131 is characterized as "severely inadequate to meet the current industry demands for distributed, flexible automation systems." Academia accepted the IEC 61499; a big number of publications have been produced and a debate on pros and cons is active [6][7]. However, industry has not accepted this standard [3].

On the other side, the IEC 61131 has been recently upgraded with a new version, i.e., version 3.0 [8], which provides support to the Object-Oriented (OO) paradigm. CoDeSys 3 [9] has already implemented an OO version of the IEC 61131 and other industrial vendors, e.g., Beckhof [10], are moving to this direction.

However, programming in an object oriented way is not a trivial task for industrial automation developers that are already accustomed with version 2.0 of IEC 61131. A long period is required to make the shift in structuring IASs in a complete OO approach. Hopefully, this transition is also pushed by new developers that enter the field since OO and UML/SysML are already in university but also in technical schools curricula for several years now. For example, OO has already become mandatory in the curriculum for technicians in Germany [11].

Except from training, specific frameworks may facilitate this transition and bring the benefits of applying OO and Model Driven Engineering [12][13] in the industrial automation domain. These challenges have already attracted the interest of academia and several works have been published towards this direction, as for example [14-16]. It is widely accepted today that the factory automation industry is slowly but steadily experiencing a paradigm shift [17].

At the same time, the traditional development process of IASs according to which the constituent parts of the systems, i.e., mechanics, electronics and software, are developed independently and then are integrated to compose the system, is criticized by many researchers, e.g., [18][19][20], as inadequate to address the always increasing complexity of these systems.

In this paper, synergistic integration at the component level, which is proposed in our previous work, is adopted as a means to address the new challenges in Mechatronic systems such as IASs. The Mechatronic component (MTC), which consists of mechanics, electronics and software, is considered as the key construct for the composition of Mechatronic Systems. We describe a cyber-physical system-based approach, which exploits composability and compositionality [23] on the MTC or cyber-physical component (CPC) level to define the system level functional and non-functional properties. The proposed approach, which utilizes synergistic integration at the cyber-physical component level, adopts the OO paradigm and exploits SysML for system level modeling and UML for modeling the software part of the system.

The main focus of this paper is on modeling the cyber part of the system and especially its software part. Two alternative implementations of the proposed design are discussed. One is based on IEC 61131 3.0 for PLCs; the other is based on a general purpose OO programming language to allow the use of the various embedded boards mainly based on ARM processors that appear in the market during last years. The whole design and the prototype implementation are in the context of an educational approach. This approach emphasizes on teaching students and industrial practitioners, the use of higher layers of abstractions and the differences of the two implementation alternatives so as: a) to successfully realize the move to the OO paradigm, and b) successfully apply the model driven development (MDD) paradigm. Moreover, a focus is given to clarify the scan cycle model that is widely used in PLC programming, since the users of embedded boards are accustomed to the event triggered programming paradigm. Our objective is for the framework to support scheduling abstraction [24], which would enable the developer to neglect the scheduling of components, and also timing abstraction, which will allow the developer to neglect timing issues and consider only causality.

The Liqueur Plant system, a case study used in [3], has been designed based on the presented approach starting from the system level. The model of the cyber part, which is derived from the system level model, is refined, following the proposed architecture, for: a) the cyber parts of cyber-physical components, and b) the process controllers, which are modeled as cyber components. For the prototype implementation, the software part was implemented using Java and the scan cycle model, and it was tested with a simulator of the physical plant. Java is considered a technology that may speed up the adoption of advances in general purpose computing in the domain of IASs [25].

The remainder of this paper is organized as follows. Section 2 refers to related work. In Section 3, the case study used in this paper is described and the basic concept of the system level based approach is briefly presented. Section 4 describes the proposed in this paper architecture for the cyber part of the system. In Section 5, two implementation alternatives are discussed and finally the paper is concluded in the last section.

## II. Related Work

Progress in general purpose computing has attracted the interest of industry and academia from the domain of IASs and several approaches to exploit the new technologies have been proposed so far. Object and service orientation, component based development and MDD are among the ones that have been extensively promoted as been able to improve the development process of IASs.

Object-orientation has already attracted the interest of the research community and various approaches have been published on how to exploit OO in this domain, e.g., [26][27]. However, as it is claimed in [28] most of these works do not take into account the OO aspects of IEC 61131 in the direction of extending it to support the OO paradigm and this has resulted to inefficient proposals regarding the OO extension of the IEC 61131 model.

Service oriented architectures (SOA) have already attracted the interest of researchers in the industrial automation domain, e.g., [17][29-32] and vendors are already moving towards exploiting the Interent of Things (IoT), e.g., [33]. To support discovery and composition of capabilities of entities that constitute CPSs and their just-in-time assembly, authors in [34] describe an approach to enable the use of SOA methods for this domain.

SysML was defined as a language for system modeling and is widely utilized in mechatronic systems modeling, e.g., [35][20], and not only for reverse engineering as claimed in [36]. Moreover, SysML is not object oriented as claimed in [36]; it may also be used for modeling systems in a procedural way.

Authors in [11] report their experience from an evaluation of a UML-based versus an IEC 61131-3-based Software Engineering Approach for teaching PLC Programming. Even though two different levels of abstraction in software specification are compared, the finding are quite interesting and can be utilized to facilitate the shift to the OO programming paradigm in IASs development. Both, UML and SysML, are utilized in the development process of IASs, e.g., [14-16][23][37]. However, most of the approaches do not use UML and SysML to increase the abstraction level in modeling so as to facilitate the development process in this domain. To our understanding, an effective use of UML and SysML should support the construction of more abstract models than the ones supported by the IEC 61131 constructs.

Component based development has been proposed by various research groups to increase flexibility and effectiveness of the development process [38-40]. Authors in [36] arbitrarily restrict the composition hierarchy in just three levels. They decompose the system into modules, and the modules into components. However, they do not argue on this, not even provide any definition of the module and the component. Moreover, it is not clear if modules or components are mono or multi discipline ones. Composability and compositionality are defined as main properties in component based development [23]. Composability ensures that, when components are properly integrated to compose a system, their properties do not change. Compositionality ensures that properties of the system can be computed from components' properties. Both properties are prerequisites for an effective component based development process.

MDD has been proposed as a paradigm to increase the effectiveness of the development process of IASs, e.g., [38][40][36]. However, it should be noted that MDD is not a tool for realizing the "integration among the different domain of mechatronic Systems" as claimed in [36].

Authors in [22] define three design layers and allocate cyber-physical Objects at the physical layer. Our approach in high-level modeling of CPSs differs in that it considers the cyber-physical component (CPC) as an integration of entities from the three layers, i.e., physical, electronics and software, and we allocate CPCs at the Mechatronic or cyber-physical layer that is on top of these three layers of the MIM architecture. In this way, the CPC encapsulates the heterogeneous interactions among the different discipline constituent parts facilitating the integration of CPCs at the system level.

Basile *et al.* describe in [3] an approach for modeling the cyber part of IASs. They propose an extension to the IEC 61131 model to support, according to authors, an event-based execution order in a similar way with the IEC 61499 standard. Authors define the structure of the cyber part consisting of two types of FBs. Device FBs (DFBs) are used to implement the basic functionality that requires access to I/O field signals. Operation FBs (OFBs) are used to implement operations that use the functionalities provided by DFBs to perform specific working cycles. To facilitate the development of distributed automation systems avoiding the disadvantages of the IEC 61499 standard, authors also adopt supervisory control to solve the problem of programming the concurrent FB behaviors so as to satisfy desired sequencing and logical constraints. Our proposal addresses several shortcomings of the above approach and may be considered complimentary to it.

Synergistic integration of the three discipline flows of Mechatronic systems, i.e., material transfer, energy conversion and information processing, is proposed by many researchers as a tool to address the always increasing complexity in the domain. The traditional development process is criticized, e.g., [18][19][20], as inadequate to address the always increasing complexity of these systems. Authors in [21] claim that the dynamic coupling between various components of a mechatronic system requires an integrated approach instead of considering the different domains separately and sequentially. System integration is considered as the elephant in the chine store of large-scale CPSs [22]. The author in [12] claims regarding the development of software that "the lack of an integrated view often forces developers to implement suboptimal solutions" and that these solutions unnecessarily duplicate code, violate key architectural principles, and complicate system evolution and quality assurance. This claim becomes stronger when applied to the system development process.

## III. A Cyber-Physical system-based approach for system development

### A. *The liqueur plant case study*

The case study used in this paper is based on the one used in [3]. We further assume that the plant is used to produce two types of Liqueur, hence we call it Liqueur Plant. Figure 1 presents the mechanical part of the plant, i.e., the physical part of the target system that performs the physical processes. The plant is composed of four silos connected by a pipe. Each silo *i* has an input valve *INi* and an output valve *OUTi* through which is cyclically filled and emptied with liquid. It also has a sensor *Ei* for the lower level and a sensor *Fi* for the upper level. Silos 2 and 4 have a resistance *Ri* to heat the liquid and a sensor *Ti* to monitor the temperature. The other two silos, i.e., silos 3 and 4, have a mixer *Mi* to mix their content.

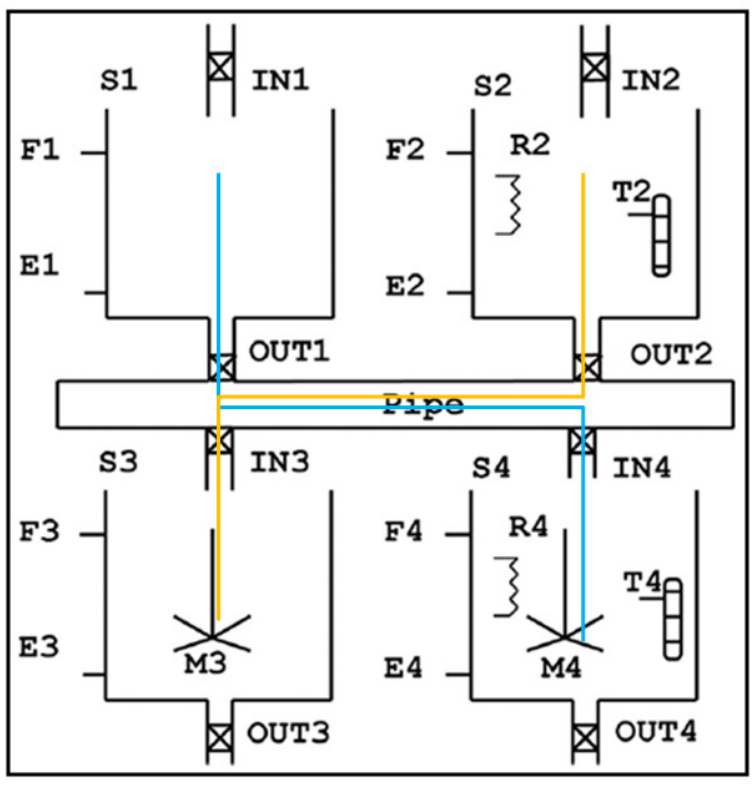

Figure 1. The physical plant used as case study in this paper [3].

Simplified descriptions of the two processes, which are executed in the plant, are assumed. Silos S1 and S4 are used for the production of liqueur of type A. Raw liquid undergoes a basic process in S1 and then it is poured into S4 where it is further processed, i.e., it is heated and then mixed. Silos S2 and S3 are used for the production of liqueur of type B. Raw liquid is heated in S2 until a given temperature is reached and then it is transferred to S3 where it is mixed for a given time. The two liqueur generation processes are independent and can be executed in parallel assuming that they use the pipe in an exclusive way. Moreover, mixing the liquid in silos S3 and S4 at the same time is not permitted due to a constraint in power consumption.

### B. *The System Level Modeling Process*

The development process we have adopted is based on refinement of models of decreasing level of abstraction as captured on the MTS-V model [35]. The SysML requirement diagram and essential use cases are used for requirements modeling. The developer captures the required at the system level functionality in terms of system responsibilities, as well as required QOS characteristics. The requirements model is next used to construct the system architecture, which defines the system as a composition of components and connectors among these, as shown in figure 2, where the proposed system level architecture for CPSs is presented. A *CyberPhysical-SystemComponent* may be *CyberPhysicalComponent*, *CyberComponent* or *PhysicalComponent,* i.e., the kinds of components that constitute a CPS. The V model, instead of what is claimed in [36], does not impose the design of the system to be separated into the development of single components, which should be designed in parallel in the single disciplines and then be integrated to the overall system.

The system architect has to split the system level functionality that includes physical processing into chunks of subsystem or component level functionalities. Chunks of system level functionality that include physical processing should be allocated to cyber-physical components, captured as *CyberPhysicalComponent in figure 2*. Chunks of system level functionality that would probably be provided by already existing CPCs are identified. For example, mixing and heating functionalities, which are required to fulfill the corresponding requirements of the raw liquid to generate Liqueur, would probably be provided by a heating and mixing Silo CPC. Based on this, the architecture of the system is defined as a composition of existing or well defined CPCs and the connectors required to interconnect them. A *CPS* usually interfaces to humans and may be considered as composed of subsystems, as shown in figure 2.

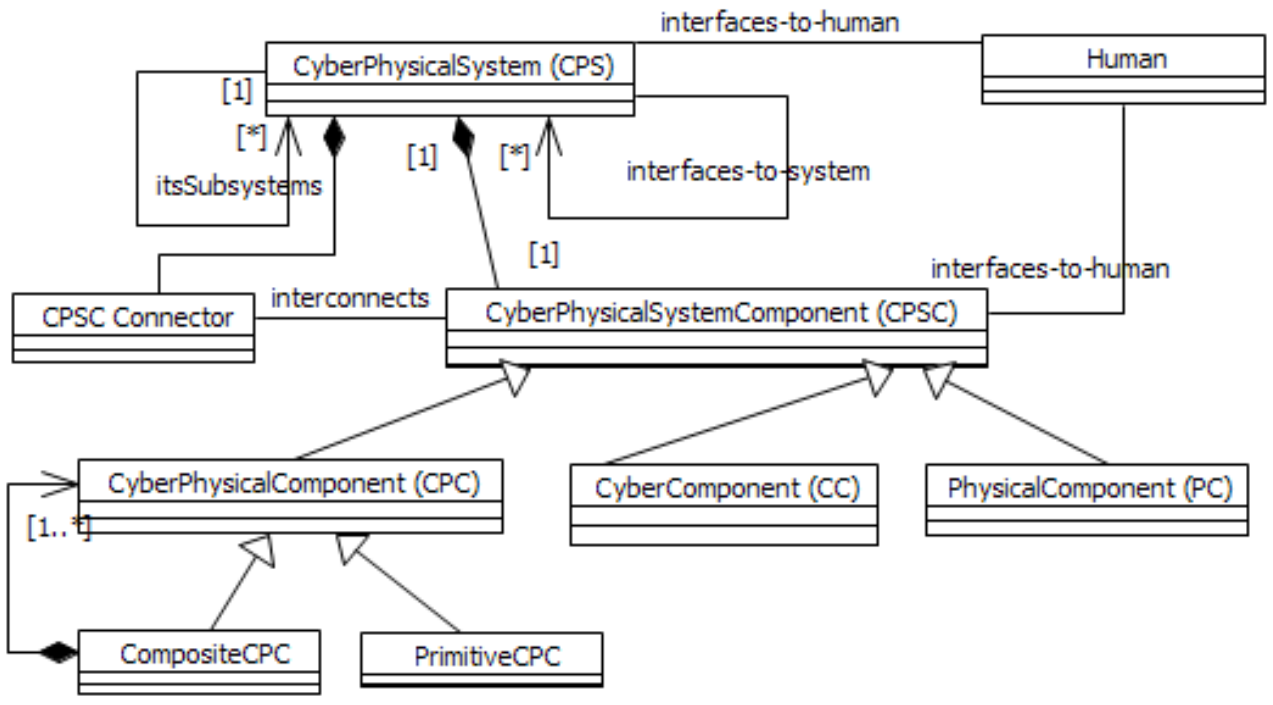

Figure 2. Proposed system level architecture for the CPS.

Hopefully, vendors of CPCs would have already developed components that provide these functionalities. They would have published in the semantic web, in a machine readable way, the information that is required for using these components. Developers may use their browsers to find the proper components [29]. They have to download the virtual CPCs, i.e., the ones that instead of the real world mechanical part contain their simulator. Using the proper development environment they integrate the virtual CPCs and construct the virtual CPS. This can be used for model analysis and verification. Afterwards, the real CPCs can be ordered and integrated to construct the target system.

Based on the above process, the Liqueur Plant is defined at the first abstraction level of the architecture model as a composition of the following CPCs: a) a simple silo (Silo), b) a heating silo (HSilo), c) a mixing silo (MSilo), d) a mixing and heating silo (MHSilo) and e) a pipe (Pipe). The refined system architecture model, which is partially shown in figure 3, captures also cyber components, which are required to capture the coordination logic of the constituent CPCs so as to fulfill the system level requirements. These cyber components may capture the coordination logic in a static or in a dynamic way. In the latter case, the functionalities offered as services at the cyber-physical component level can be orchestrated to define the required system level functionality. For example, the specific component collaboration or orchestration of services to generate liqueur of type A represents system specific logic, and is captured by the cyber component *GenLiqueurA*, which utilizes services offered by silo 1 and silo 4 as shown in figure 3.

The Pipe has been represented as a cyber-physical component that is defined as a specialization of the cyber component CommonResource, which captures the logic of acquiring and releasing a common resource since it was decided to implement this logic by software. Common structure and behavior of the various processes has been captured in the LiqueurProcess cyber component (see *CyberComponent* in figure 2) that will also be inherited by the GenLiqueurB cyber component (not shown in figure 3), which captures the coordination logic of generating liqueur of type B. The cyber component *PlantController* captures coordination logic of the various plant processes.

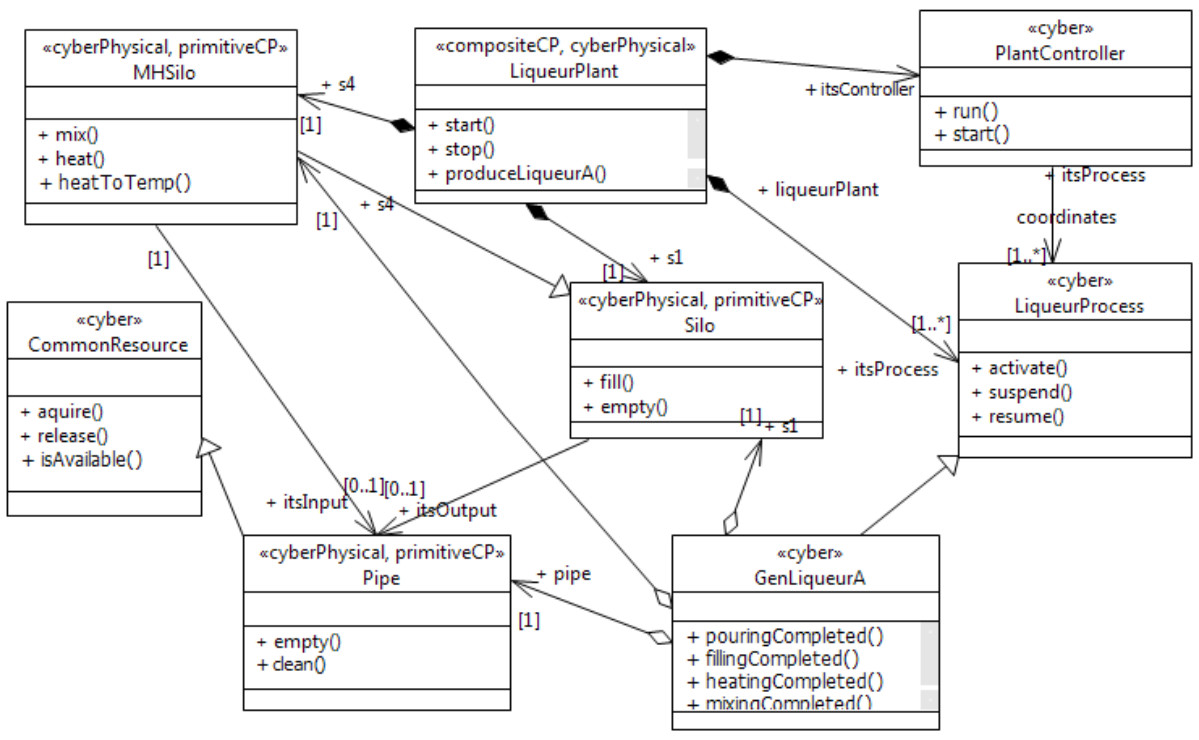

Figure 3. Part of the architecture of the LiqueurPlant system.

## C. *The proposed CPC architecture*

Figure 4 presents the proposed architecture for the CPC using the MHSilo CPC as example. The CPC is composed of the physical part, i.e., *itsPhUnit* and the cyber part, i.e., *itsCyberPart*. The cyber part of the CPC is further decomposed into: a) the software part (*itsS-part*), which represents the software required to transform the physical unit into a smart unit, and b) the electronic part (*itsE-part*), which represents the processing node required to execute the software part. As described in the next section, the software part is further decomposed into the software representative of the physical unit into the software domain (*itsSR*), and the controller (*itsController*).

SysML ports [41] are used to represent the interaction points of the CPC with its environment. The type of the port specifies features available to and requested from the external entities via connectors to the port. For example, the *itsProcessPort* of *MHSilo* is of type *ProcessPort*, which specifies the features of its internal part *itsController* that are visible through external connectors to the environment of the *MHSilo* CPC. It should be noted that the *processPort* is of type *proxy port* since it acts as a proxy of the *itsProcessPort* of *itsControler*. On the other side the *itsProcessPort* on the border of the *itsControler* part is a *full port* since it defines with its own features the interaction point in its boundary. The implementation of the port design space construct is described in the next section. Figure 4 does not capture the interactions between the constituent parts of the CPC.

Traditional communication technologies in industrial automation can be utilized for the integration of cyber and cyber-physical components. However, in order to bring into the industrial automation domain the benefits of IoT, we adopt for the integration of cyber and cyber-physical components the Intranet of Things.

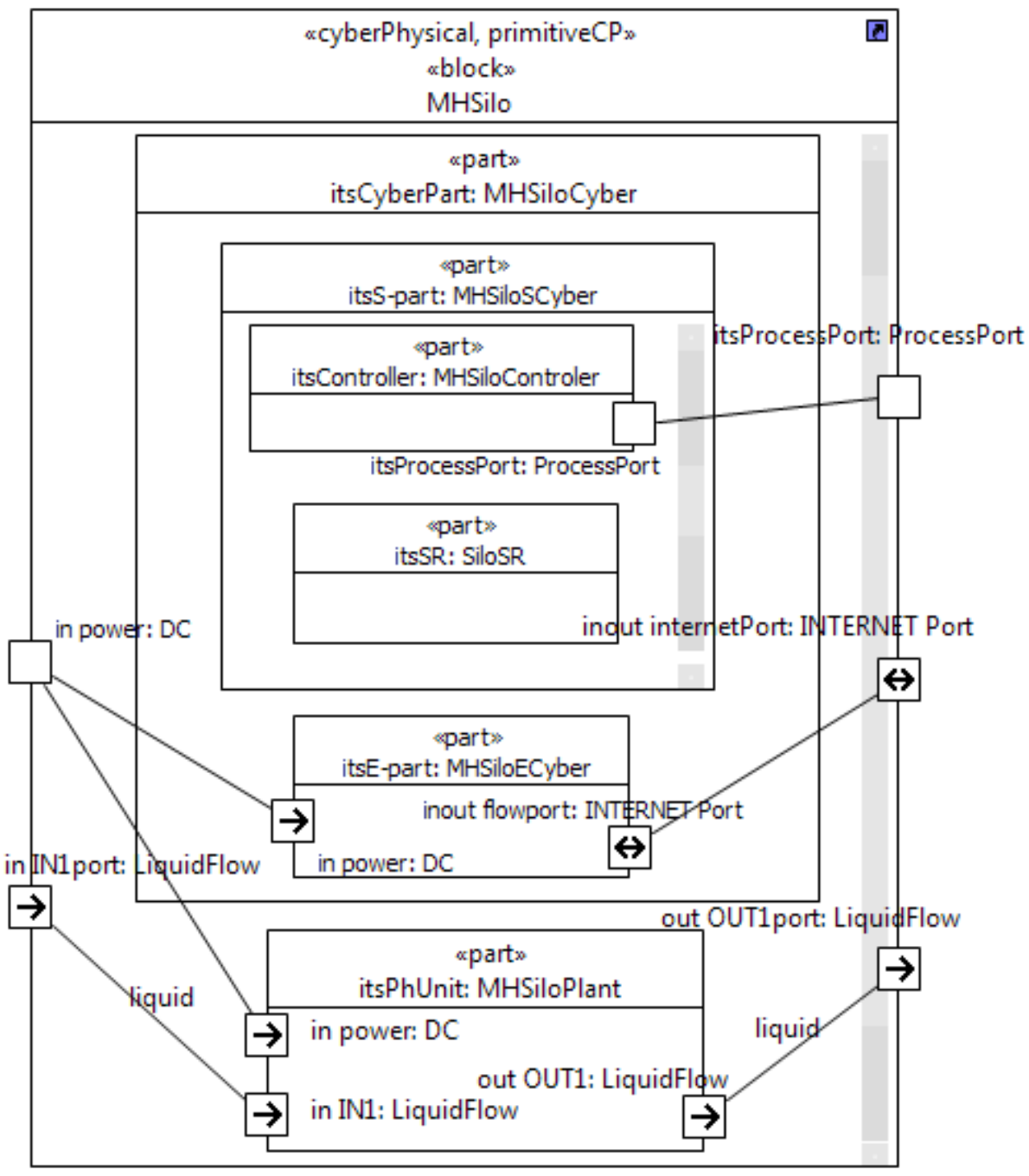

Figure 4. Architecture of cyber-physical component. Ports define the interaction points of the CPC with its environment.

## IV. Design Of the Cyber Part

The described in this paper design approach for the cyber part of the system can also be applied in the traditional development process, where the software is considered as a whole for the system. In this case, the software part is either developed synchronously with the development of the mechanics and electronics or it may be developed when these parts have already been developed and integrated. In this section, we firstly describe the application of the proposed design for the multidiscipline-component approach adopted in this paper and then for the traditional development process.

### *A. The system as a Composition of Cyber-Physical Components*

In the case that the system is considered as a composition of cyber-physical components the first design of the cyber part of the system results from the SysML-view model. This constitutes the cyber-view of the system and is generated by projecting the system model, which is expressed in SysML, to the cyber domain. The cyber-view can be considered as the software view (S-view) of the system, since: a) we have decided to assign no chunks of functionality to electronic parts but assign the whole control functionality to software and, b) only system level functionality is captured in the system model at this stage of modeling. The trend today is to assign all the functionality of the controller to the software part and use general purpose embedded boards or PLCs just for the execution of the software part. We adopt this approach and as a result we do not capture the electronic part of the cyber-physical system at these early design documents.

A cyber component is generated in the cyber-view, for every cyber-physical component of the system level model. This cyber component is assigned the behavior of controlling the physical component to perform the services that have been defined at the component level. Thus the MHSilo cyber component has been assigned the behavior of controlling the MHSilo physical component to perform the operations fill(), empty(), heatToTemp() and mix(), which have been assigned to the corresponding *MHSilo* cyber-physical component. Moreover, every cyber component at the SysML view results to a component at the cyber view. The GenLiqueurA component is an example of such a component.

The so created architecture is further refined so as to be implementable in an effective way. Modularity and reuse are primary concerns in defining our proposal for further refining the software part of the automation system. Specific UML design constructs if properly used will provide a solid framework for semi-automating the refinement process but also provide the information required for automating the generation process of the implementation code. The software component of every cyber-physical component of the system level is further decomposed into: a) the software representative (*SR*) and b) the controller.

The *SR* is the representative of the corresponding physical part in the software domain. It acts as the proxy of the real world object in the software domain and encapsulates all the interface details with the real world object. It should be noted that the SR does not add any further functionality to the one provided by the real world object. It captures not only properties but also information of the real world object that may be of interest to the software part of the system, such as model type, manufacturer, serial number, dimensions, etc. For example, the *itsSR* of type *SiloSR* in figure 4 is the representative of the physical Silo in the software domain.

The *controller* captures the logic required to convert the low level operations performed by the physical part to more sophisticated operations offered at the CPC level. The *itsController* in figure 4 is an example of such a component. The *MHSiloController* may be used to implement operations such as fill(), empty(), mix() and heatToTemp(). In other words the controller transforms the physical object to a smart object. Different controllers may convert the same physical object to a different smart object with the controller to play a key role in this physical object transformation process.

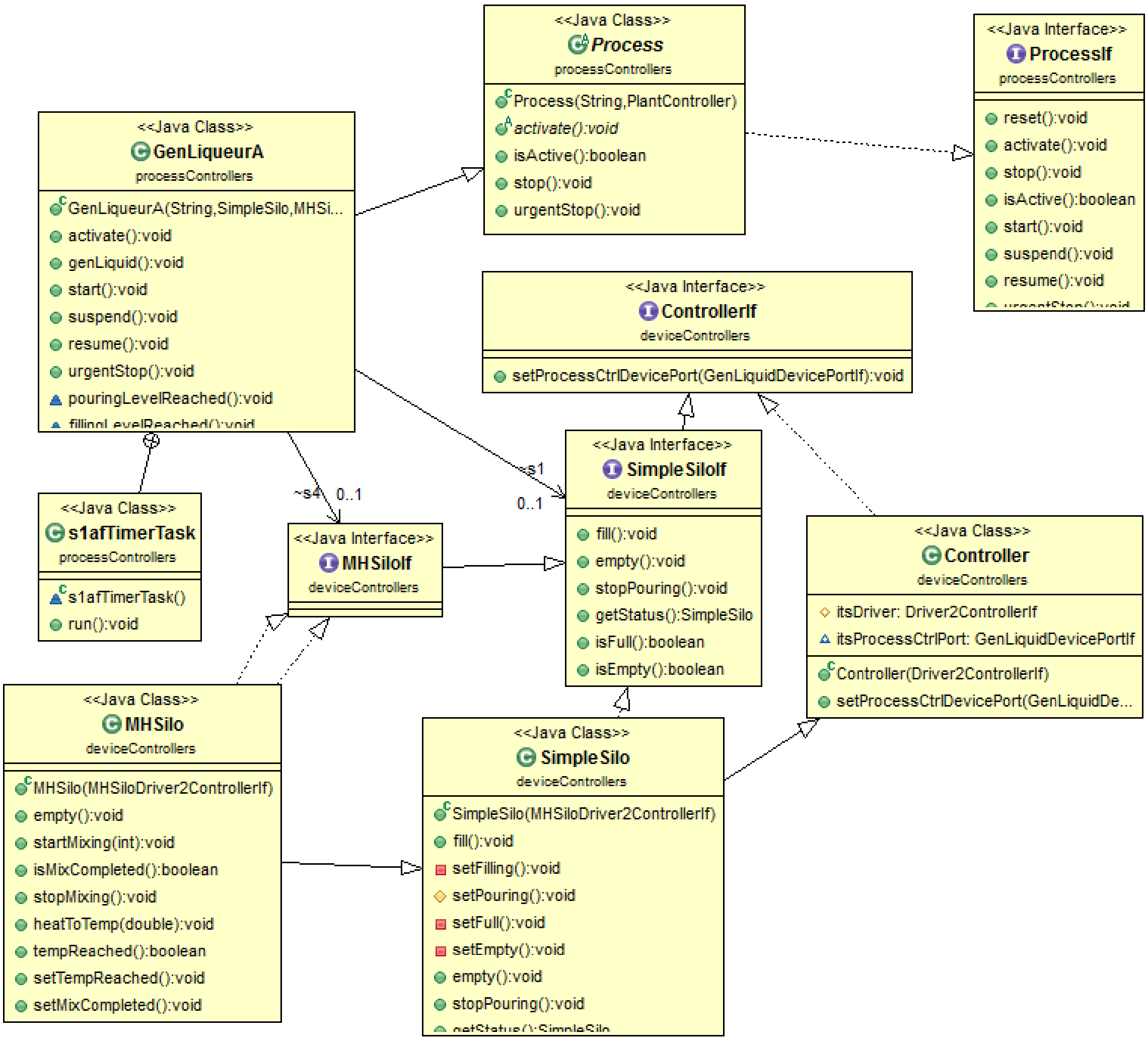

Figure 5. Part of the architecture of the Cyber part of the Liqueur Plant system using predefined classes.

In figure 5, a part of the cyber-view of the system is given. This part captures the structure of the software that is used to realize the process of generating liqueur of type A. As shown in figure 5, the cyber components are represented only by their software constituents. However, in case that the developer decides to assign chunks of system level functionality to electronic components then these components should appear in the cyber-view of the system at this level of specification.

There are two alternatives to capture the basic structure of the proposed framework. One is through predefined generic classes that capture this knowledge and allow the developer tore-use it through the mechanism of generalization-specialization. In figure 5, which is based on this approach, classes *Process* and *Controller* as well as interfaces *ProcessIf* and *ControllerIf* represent predefined design constructs used to capture the basic structure adopted by the framework. These classes are used by the developer to define the classes of her system. Thus, *GenLiqueurA* is defined to inherit the *Process* class and the *SipleSiloIf* is defined to inherit the interface *ControllerIf*. The use of interfaces allows an independent of silos' implementations definition of the *GenLiqueurA* class. Based on this, the *GenLiqueurA* class may be used with any cyber-physical silo that implements the specific interfaces. On the upper level, the use of *ProcessIf* imposes a low coupling between the process and the *PlantController* cyber component.

Based on the second alternative for reusing the framework's knowledge, a profile is used to define specific stereotypes such as *«controller»*, *«controllerIf»*, *«process»* and *«processIf»*. Figure 6 presents part of the architecture of the cyber part of the Liqueur Plant system exploiting the appropriate profile. It is evident that the profile mechanism simplifies the design specification and makes the design process more friendly to the developer.

### *B. The system as a composition of Cyber and Physical parts*

In the case of the traditional development process of Mechatronic Systems the software is considered after the development of the hardware parts. In this case, the structure of the already developed part of the system is used as a basis for the definition of the structure of the software. In the high level of abstraction of the software architecture there is a software component for every physical unit and also for every plant process. It is evident that the constraints in defining the software level in this case are more, compared to the 3+1 SysML-view model approach since the implemented software has now to be compliant with the already implemented hardware which defines the infrastructure on which the system services have to be implemented. This approach can be characterized as a middle-out approach since the implemented software has to meet requirements imposed not only from the system layer specification but also from already defined lower layer services that the software has to utilize in order to meet system requirements. On the other side, in the top-down approach, the one applied when the system is considered as a composition of cyber-physical components, the software is concurrently and synergistically developed with electronics and mechanics to meet the requirements of the system level. This approach does not impose constraints on the software solution space as is the case with the middle-out approach.

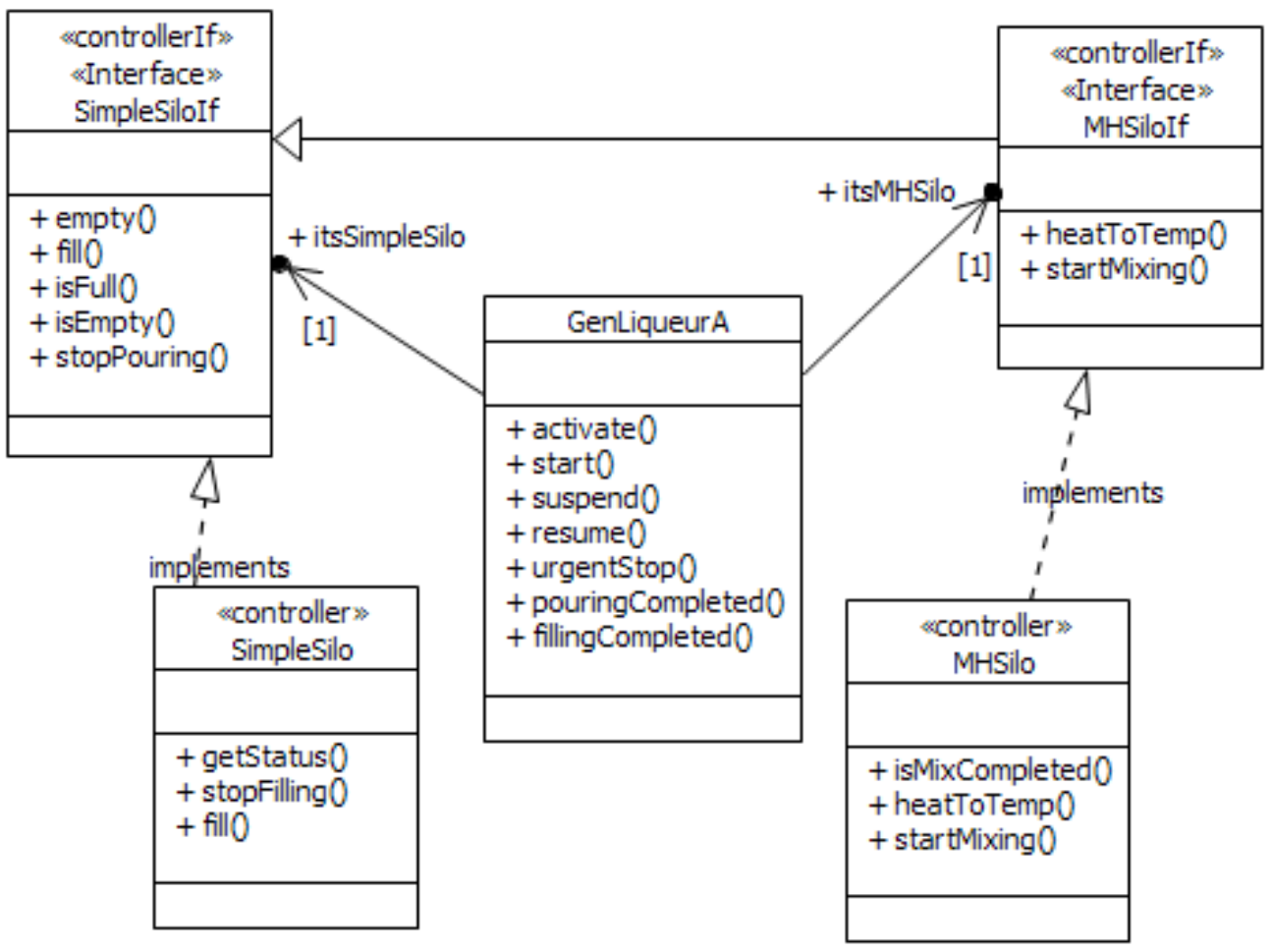

Figure 6. Architecture (part) of the Cyber part of the Liqueur Plant system based on a UML profile.

## V. Implementation Alternatives

Industrial automation software is typically implemented based on the scan cycle model. According to this model, plant inputs are read, controller code is executed based on these inputs and the generated outputs are written to the plant to affect its operation. This is known as READ-EXECUTE-WRITE cycle and is an implementation of the time triggered control. The cycle time has to be properly defined to address the response time requirements imposed by the physical plant.

In this section, we present two implementations for the proposed in this paper design. One based on the widely used IEC 61131 standard and more specifically on version 3.0 of the standard that supports object-oriented programming and the other on a general purpose object oriented language. Java was selected as language of this type. Both implementations are based on the scan cycle model. A key issue in these implementations is that the framework has embedded the high level control flow of the application so the developer has to focus on the selection of components that will constitute the system and their integration. This greatly simplifies the construction of this kind of systems. The high level control flow is that introduces several bugs and makes the programming of these systems difficult to comprehend.

### A. *A Java based implementation*

The proposed design can be implemented in Java and executed on the various embedded boards, mainly based on ARM processors, that recently appeared in the market. A mapping to Java constructs is proposed in this section so as to apply the scan cycle based execution. The design of the software part of the cyber-physical system has been implemented in Java and executed with a simulator of the liqueur plant to prove the effectiveness of the design. The system has been developed as educational tool to be used in labs in related courses. An executable version of this implementation is available for download from the web site of the author.

Figure 7 presents a snapshot of this implementation, where Silos S1, S2 and S4 have been instantiated. Using this system, students may comprehend the differences of a traditional Java based implementation of the plant controller from the cycle based one and identify challenges and technologies that provide solutions to both implementations. They can be exposed to the use of higher layers of abstraction and also to the various mechanisms of the implementation environment to address mutual exclusion problems that appear in the traditional Java event-based implementation and compare this with the time triggered implementation that is adopted in PLC systems. The scan cycle based Java implementation is based on the use of Java Timers. This is actually the implementation of the Task construct of the IEC 61131 model.

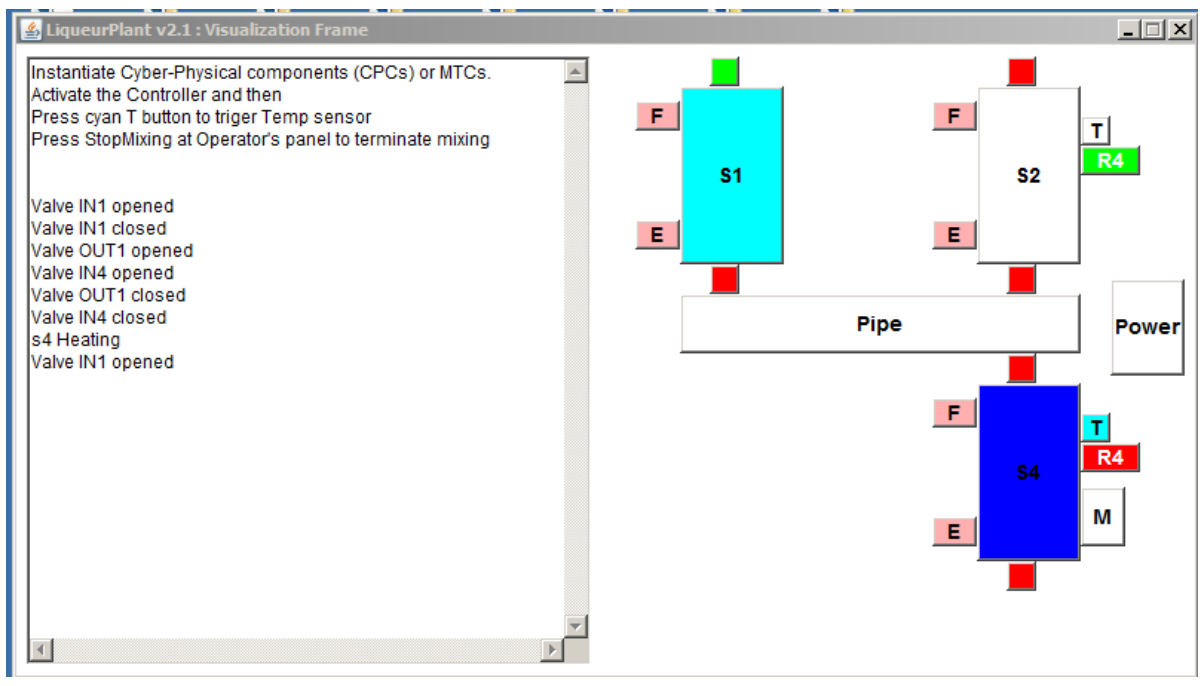

Figure 7. Snapshot of the educational version of the Liqueur Plant example application.

The LiqueurPlantSystem in this implementation is defined as a composition of CPCs. A visualization frame is used to instantiate the required virtual CPCs and the plant processes such as *GenLiqueurA* and *GenLiqueurB*. Power has been modeled as resource to re-use the already developed generic infrastructure for handling the pipe as a common resource. The PlantController setups the timer and defines through the execute operation of the `controllerTimer Task` the activity that should be executed by this timer, as following :

```
controllerTimer = new Timer();
controllerTimer.schedule(new
  controllerTimerTask(), 0, 500);
```

where the controller has been defined to extend the TimerTask and so override its run method.

## *B. IEC 61131 based implementation*

The new version of IEC 61131, i.e., version 3.0, which provides OO support, allows for a more straightforward mapping of the SysML/UML design specs to the implementation language constructs. However, a mapping to the widely used today IEC 61131 is also possible, exploiting the already existing OO support that is provided by version 2.0 of the standard. An example of such an implementation is given in [3].

The FB and the class constructs of the IEC 61131 are used to implement classes of the design specification, while the interface construct is used to implement interfaces. Figure 8 presents example IEC 61131 code that will be automatically generated from the design specification.

Cyber components of type *«process»* can be hosted in cyber-physical components that offer hosting services of the required QoS, if any, or they may be allocated to shared or exclusively used execution environments. In the remaining of this section the implementation of the port design construct is presented.

```
FUNCTION_BLOCK MHSILO_CONTROLLER
....
itsPROCESS_PORT:MHSILO_PROCESS_PORT;
itsDRIVER_PORT: MHSILO2DRIVER_PORT;
....
END_FUNCTION_BLOCK

FUNCTION_BLOCK MHSILO_PROCESS_PORT EXTENDS
CONTROLLER2PROCESS_PORT IMPLEMENTS MHSILO_IF
...
itsPROCESS:PROCESS2MHSILO_IF
...
END_FUNCTION_BLOCK

INTERFACE PROCESS2MHSILO_IF
METHOD FILLING_COMPLETED END_METHOD
METHOD POURING_COMPLETED END_METHOD
METHOD HEATING_COMPLETED END_METHOD
METHOD MIXING_COMPLETED END_METHOD
END_INTERFACE
```

Figure 8. Sample IEC 61131 code.

The proposed implementation for the port construct provides very low coupling between the system components. Figure 9 presents how the two ports of the *MHSilo* controller, which is shown in figure 4, are implemented based on the proposed approach.

One port, the *itsProcessPort*, which is used to interconnect the controller with its process, is of type *MHSiloProcessPort,* which inherits the *Controller2-ProcessPort* and implements the *MHSiloIf.* This port contains also the data member *itsProcess* through which the *MHSilo* component accesses the services that it requires from the process level corresponding component that will be interconnected to this port. These required services are specified by the type of the *itsProcess* data member that is of type *Process2MHSiloIf* as shown in figure 10.

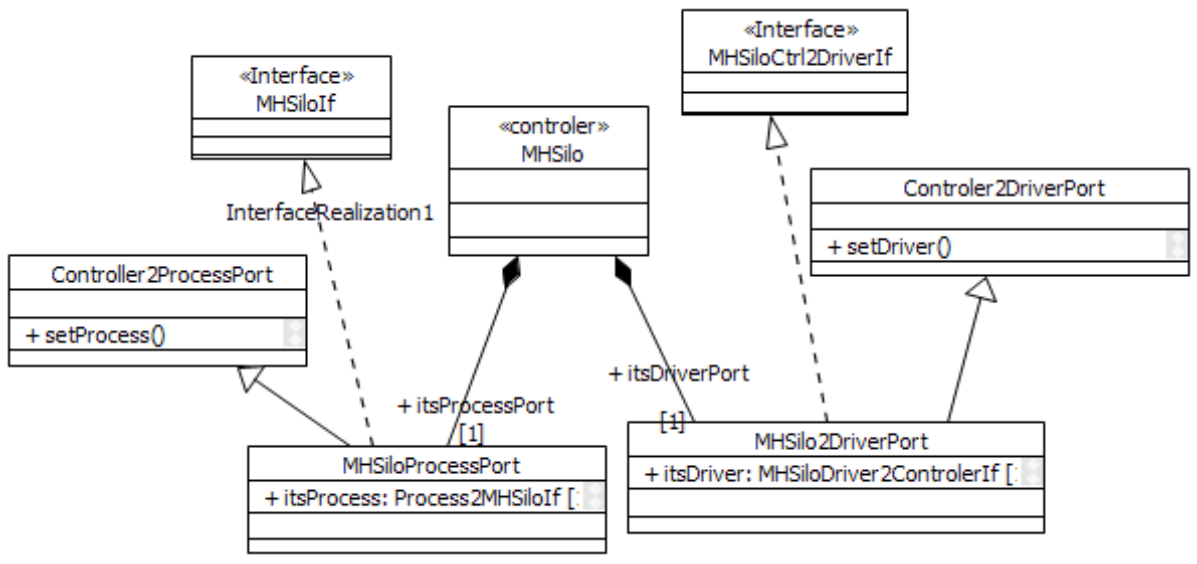

Figure 9. Implementing ports for the MHSilo Controller.

The other port, the *itsDriverPort,* which is not shown in figure 4, is used to interconnect the controller with the corresponding SR in the case that a higher modularity is also required for the cyber part of the cyber-physical component. This port implements the *MHSiloCtrl2DriverIf* and includes a data member *itsDriver* of type *MHSiloDriver2ControllerIf* (see figure 11). It should be noted that due to high coupling between the physical part and its SR the first level of reusability is the one above the SR. However, also at this level the coupling between the SR and the controller is high in most of the cases. Due to this high coupling between the controller and the SR that results to a low potential for reusability at this level, interfacing with ports at this level is considered not effective. This is also the reason for considering as effective reusability the CPC level and adopt a more effective compared to port implementation of interfacing between SR and the controller.

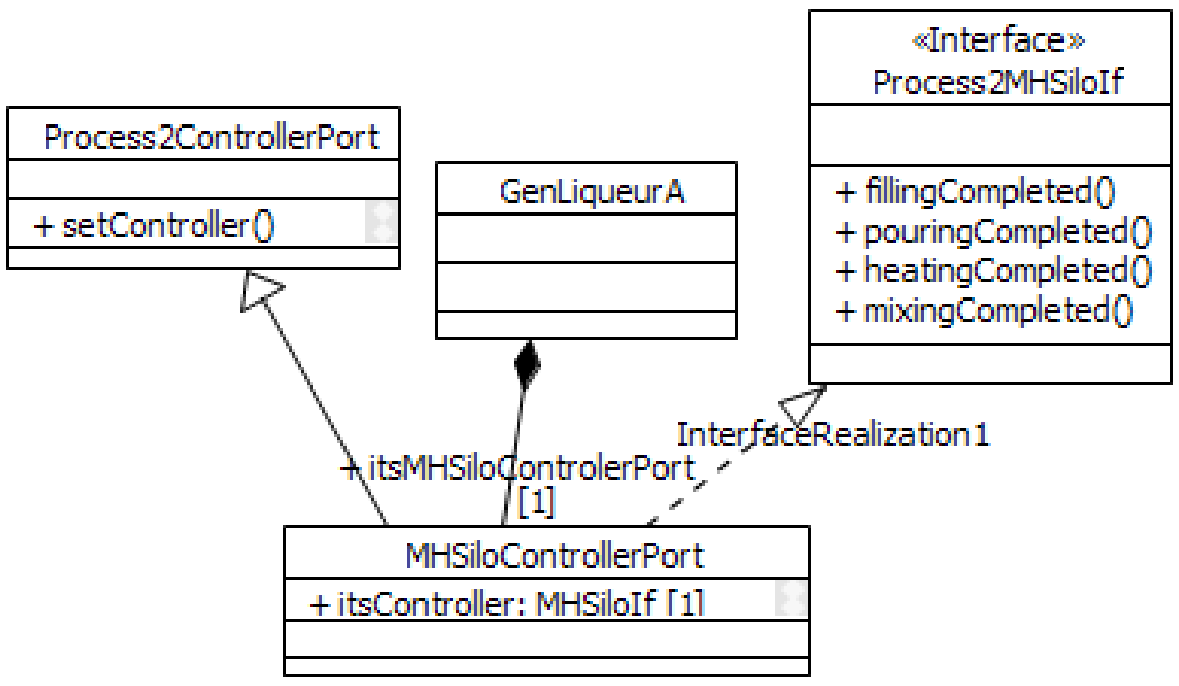

Figure 10. Implementing ports for the GenLiqueurA Process.

Compliant ports may be interconnected by means of connectors. Establishing a connector means to setup the corresponding data members of the connected ports.

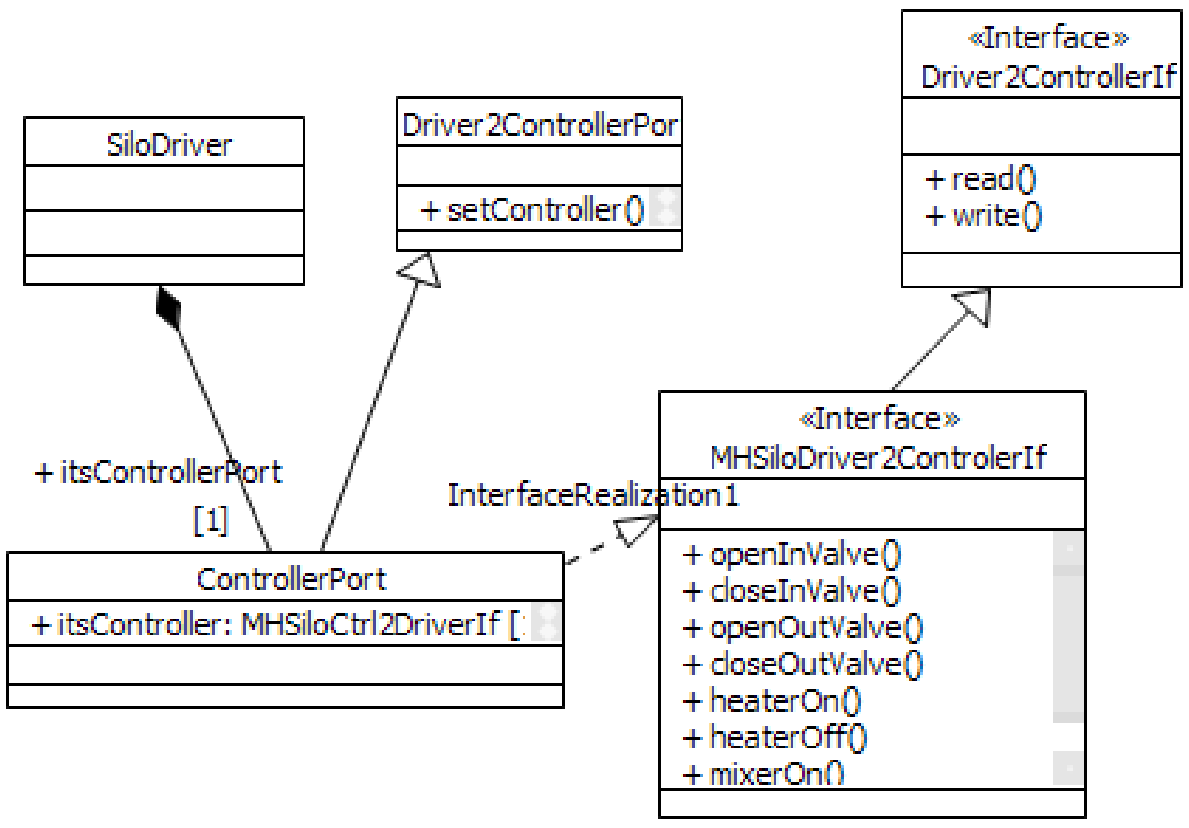

Figure 11. Implementing ports for the Silo Driver.

## Conclusion

The shift from the traditional IEC 61131 and PLC based development process to the new OO version of the IEC 61131 is not an easy task for the industrial developers, who are familiar with this programming paradigm for many years. Knowledge of the basic concepts of the OO paradigm is required, but also in this case the shift is quite complicated. A framework that greatly simplifies this process has been presented in this paper. This framework is based on a cyber-physical system-based approach and can also be used to generate implementations on the various ARM-based embedded boards that have recently appeared in the market. An example application has been developed in the form of a lab exercise to be used by students and industrial developers to understand the use of higher layers of abstraction in system development and their realization using IEC 61131. The framework maybe criticized as introducing performance overhead but it greatly simplifies the development process and increases the quality of the generated code.

## Acknowledgments

Part of this work has been funded by Bayerische Forschungsstiftung in the context of an international collaboration.